\documentclass[12pt]{article}
\usepackage{graphicx}
\usepackage[english]{babel}
\usepackage[latin1]{inputenc}
\usepackage{amsmath,amssymb}
\usepackage{bm}
\date{}
\begin{document}

\title{\bf EVOLUTION OF LOCAL STRUCTURES \\ IN ALKALI-BORATE GLASSES}
\maketitle
\author{Richard Kerner$^*$, \, Dina Maria dos Santos-Loff$^{**}$  and Ana Cristina Rosa$^{**}$} \\

$*$  {Address : 
LPTMC, Sorbonne-Universit\'e - CNRS UMR 7600 , \\ Tour 13-23, 5-\`eme , Boite 121,
4 Place Jussieu, 75005 Paris, France \\
Tel.: +33 1 44 27 72 98, \,  Fax: +33 1 44 27 51 00, \\  {\small email : richard@kerner@sorbonne-universite.fr}}

$**$ {Address: Departmento de Matem\`atica, Universidade de Coimbra,  \\
3000 Coimbra, Portugal (retired). \\
{\small e-mail dina@mat.uc.pt, \hskip 0.3cm e-mail cristina@mat.uc.pt}

{\abstract{We analyze the dependence of relative proportion of various characteristic
clusters in binary alcali-borate glasses on modifier's concentration $x$.
A pure $B_2O_3$ glass contains a huge amount of boroxol rings and some amount of 
boron atoms in between, linking the boroxol rings via oxygen bonds. The addition of
the $Na_2O$ modifier creates four-coordinated borons, but the resulting network
glass remains totally connected. We study local transformations that lead to 
creation of new configurations like {\it tetraborates, pentaborates, diborates,} etc.,
and set forth a non-linear differential system similar to the Lotka-Volterra model.
The resulting density curves of various local confugurations as functions of $x$ are
obtained. Then the average rigidity is evaluated, enabling us to compute the glass
transition temperature $T_g(x)$ for a given value of $x$}}

\section{Introduction}

\vskip 0.3cm
In this article we present a  structural analysis of amorphous network formed by the boron oxide $B_2O_3$
 with addition of an alkali modifier $Na_2 O$.
 Our approach is exploiting two essential features of this network:
 {\it connectivity} and {\it rigidity}.
Both characteristics can be given a quantitative treatment, and serve as essential parameters determining 
physical properties of alkali-borate glass $(B_2O_3)_{(1-x)} \, (Na_2 O)_x$.  

The alkali borate glasses, $(1-x) B_2O_3 + x Na_2O$ or $(1-x) B_2O_3 + x Li_2O$ are well known and
present an interesting field for theoretical modelling. 
 Typical {\it structural glasses}, their physical properties are determined
by topological and geometrical features characterizing medium-range order. Therefore the statistical approach,
 dealing with {\it probabilities} of finding a given local configuration, remains the most
 appropriate methodology.
\noindent
Mathematical models exploring structural properties 
of alkali-borate glasses were initiated since the early eighties (\cite{Bray1985}, \cite{RKJNCS}, \cite{BarrioGaleener})
The most important development resulted from the collaboration with the experimentalists.
M. Balkanski and M. Massot (\cite{Balkanski})
The final version of the model, including the stochastic matrix approach, was elaborated by R.K and D.M. dos Santos-Loff, 
with R.A. Barrio, M. Micoulaut, G. G. Naumis and J.-P. Duruisseau (see e.g. \cite{RKRBJPD}, \cite{DMDSRKMM}, \cite{BarrioNaumis}).

In those models we used the stochastic matrix method combined with analysis of energy costs of forming
particular local configurations. The entropy was accounted for via average connectivity and 
coordination number. The model was focussed on glass transition temperature $T_g$ and gave fair predictions
concerning the dependence of $T_g$ on modifier concentration in alkali-borate glasses in particular (\cite{Kerner1995}, \cite{RKMM1997})
The present paper deals with a more detailed analysis of stable local structures appearing in alkali borate
glasses, whose presence was corroborated by many data obtained via Raman or NMR spectroscopy. 
The ``breathing mode" identified as the $808^{-1} $ cm line in the Raman spectrum is a firmly identified signature of boroxol rings 
(\cite{KroghMoe}, \cite{Bray1985}, \cite{Hannon1994}, \cite{BarrioGaleener}).

Here we derive a system of differential equations of Lotka-Volterra type, describing
the evolution of probabilities of various local structures as functions of alkali modifier's molar concentration. This
enables us to evaluate the average rigidity defect $<r>$ for a given alkali concentration $x$, and introduce
a simple model of glass transition temperature dependence on $<r> = <r> (x)$ akin to the Gibbs-Di Marzio formula (see \cite{DiMarzio1958}).

\section{Local structures in amorphous $B_2O_3$}

A pure $B_2O_3$ glass is an amorphous solid in which atoms form a typical random network with covalent bonds.
However, taking into account that boron atoms are three-coordinated, and that all oxygen atoms
form bonds between them, one cannot produce a random network in three dimensions without forming rings. In both
crystalline and amorphous $SiO_2$ the network contains mostly the $6$-folded rings; in amorphous $B_2O_3$ the
$3$-fold rings dominate (see e.g.\cite{KroghMoe}, \cite{Bray1980}, \cite{Walrafen}, \cite{GaleenerBarrio}, \cite{BarrioGaleener}, \cite{Kerner1995}).

In the case of the $B_2O_3$ glass we admit that the boroxol rings are the
most common structure, which does not exclude the existence of much bigger closed cuircuits, which however do not have any
particular physical signature. The ``breathing mode" identified as the $808^{-1} $ cm line in the Raman spectrum is a firmly 
identified signature of boroxol rings (\cite{Bray1985}, \cite{Hannon1994}).

The structure of random $B_2O_3$ network must therefore contain lots of boroxol rings 
as well as certain amount of isolated boron tripods $B(O_{\frac{1}{2}})_3$:
\begin{figure}[hbt]
\centering 
\includegraphics[width=5.3cm, height=4.3cm]{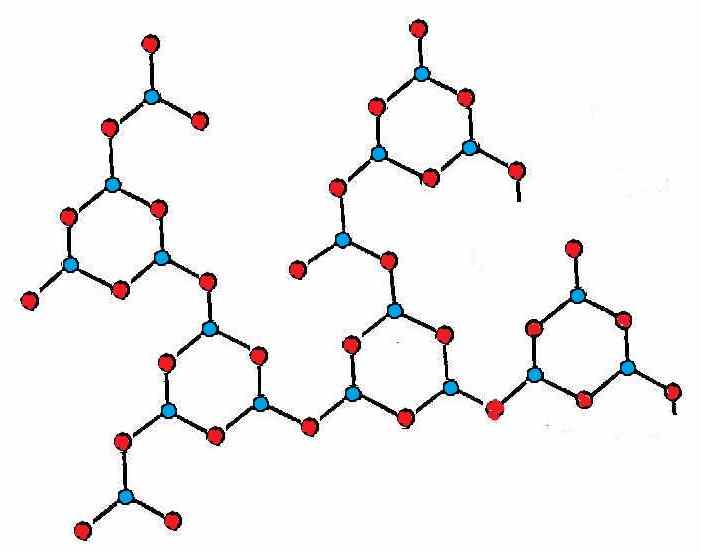}
\hskip 0.4cm 
\includegraphics[width=5.3cm, height=4.3cm]{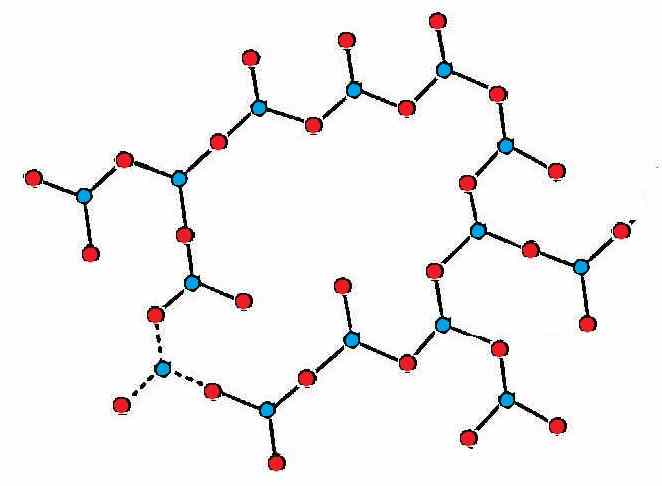} 
\caption{{\small A fragment of random $B_2 O_3$ network: Left: with boroxol rings, Right: without boroxol rings. 
Boron atoms are blue, oxygens are red.}}
\label{fig:Boreclusters}
\end{figure}

Larger units should include boroxol rings. Our aim is to include medium-range configurations
containing enough information about the boroxol ring content of the network.
There were a lot of discussions concerning short-range local structure of amorphous $B_2O_3$.
The most controversial point was the very existence and possible rate of the {\it boroxol rings},
represented in (\ref{fig:Boreclusters}). 

The relative abundance of boroxol rings has been the subject of many discussions. Certain estimates, both theoretical 
and experimental, including computer simulations, converge to the value $f \simeq 0.75$ for the relative part of boron atoms 
belonging to boroxol rings versus the total number of boron atoms in the structure. (\cite{Ferlat2008})
The lower limit seems to be that of $f=0.5$, as advocated by Wright and Vedishcheva in \cite{Wright2016}

Some authors claim an even more massive
presence of boroxol rings, corresponding to $f \simeq 0.83$ (\cite{RKRBJPD}, \cite{DMDSRKMM}), which may be considered 
the upper limit for $f$.
The models with low values of $f$ seem to be gradually dismissed by most of the authors. In what follows,
we will admit the value of $f = 0.83$, conformal to our model in (\cite{RKRBJPD} and \cite{Hannon1994}) i.e. slighly more than $80 \%$ 
of all boron atoms in pure $B_2O_3$ glass are contained in boroxol rings. Of course, this supposes an ideal case, when the glass 
is annealed very slowly letting the most homogeneous and energetically and entropically privileged configuration to be realized;
rapid quenching may lead to totally different results, similar to a frozen structureless liquid. 

The entire network can be subdivided now into $N_R$ boroxol rings $R$ and some number of isolated
borons, $N_I$. The relative probabilities to find a ring or an isolated boron are therefore
\begin{equation}
p_R = \frac{N_R}{N_R + N_I} \; \; \; {\rm and} \; \; \; p_I = \frac{N_I}{N_R + N_I}, \; \; \; p_R + p_I = 1.
\label{firstprobs}
\end{equation}
Let us denote the part of boron atoms contained in boroxol rings by $\xi$; this parameter can be expressed by $p_R$
via simple relation:
\begin{equation}
\xi = \frac{3 p_R}{3 p_R + p_I}, \; \; \; p_R = \frac{\xi}{3 - 2 \xi}.
\label{xipR}
\end{equation} 
For example, $p_R = p_I = 1/2$ as supposed by Wright and Vedishcheva in \cite{Wright2016},
will correspond to $\xi = 0.75$, which is the value advocated by Ferlat et al. (see \cite{Ferlat2008}). 
However, with such a high amount of ``free'' borons non contained in boroxol rings chains of two or even three
such entities woud appear, conveying extra floppiness and inhomogeneity to the network. In our model based
on the stochastic agglomeration (see \cite{RKJNCS}, \cite{RKRBJPD} and \cite{DMDSRKMM}) and using the binding energies 
conjectured in (\cite{KroghMoe}), we arrive to the value $83 \%$ of boron atoms trapped in boroxol rings, too.

\begin{figure}[hbt]
\centering 
\includegraphics[width=2.5cm, height=1.7cm]{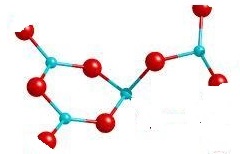}
\hskip 0.8cm 
\includegraphics[width=3.3cm, height=1.8cm]{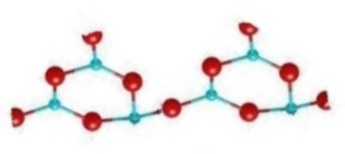} 
\caption{{\small Two typical clusters, named A and B, made with
boroxol rings and free boron tripods. Note that they have the same exterior connectivity: 4.}}
\label{fig:UnitsAB}
\end{figure}

It is easy to derive the relative frequency of $A$ and $B$ configurations
in an amorphous $B_2O_3$ network. Let the probability to find an 
$A$-configuration be $p_A$, and that of the $B$ configuration $(1-p_A)$.
Then we should have:
\begin{equation}
\frac{3}{4} \, p_A + (1-p_A) = 0.83,
\label{frac34}
\end{equation}
which yields the result $ p_A = 0.68$,
close enough to $66,67 \%$ to enable us to use the simplest
fraction available, and assume the presence of $2/3$ of clusters $A$ and $1/3$ of clusters $B$ in the network.
By the way, replacing $0.68$ by $2/3$ in \ref{frac34} would change the value $0.83$ into $\frac{3}{4} \cdot \frac{2}{3} + \frac{1}{3} = 0.83333$,
a negligible variation, below the experimental  precision.

\section{Connectivity and rigidity}

The chemical bonds in $B_2 O_3$ glass belong to the category of {\it covalent bonds}, whereas
the $Na$ atoms are linked to oxygen atoms via {\it ionic} bonding (\cite{KroghMoe1963}, \cite{Bray1985}). In binary $(1-x) B_2O_3 + x Na_2O$
glasses local structures are made of atoms strongly connected through covalent bonds, 
and ions $Na^+$ loosely connected to the main network when they stick to boron atoms, creating extra
oxygen bond, or more strongly connected to oxygen atoms, breaking bonds between neighboring borons
and creating dangling oxygen bonds. The first tendency prevails at low concetrations, while the
second possibility is realized at molar $Na_2O$ concentrations higher than $35 \%$.

Considering the network just as a collection of isolated atoms $B$, $O$ and $Na$ does not
give any structural information beyond its pure chemical content. The subdivision of random
network into molecules still does not give enough information, in particular, about the local ring
structure.  

Adding alkali modifier creates new local configurations. The molecules of $Na_2O$ dissociate
in the hot melt, and the $Na^+$ ions come close to the boron atoms of the network, thus creating
a new oxygen bond and transforming three-coordinate boron atoms into four-coordinate ones, at least
at concentrations not higher than $25 \%$ as shown in tne following figure:
\begin{figure}[hbt!!]
\centering 
\includegraphics[width=1cm, height=1cm]{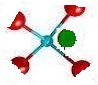}
\hskip 1.4cm
\includegraphics[width=2.4cm, height=1.8cm]{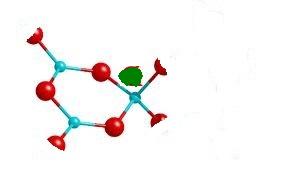} 
\hskip 2cm
\includegraphics[width=1.2cm, height=1.2cm]{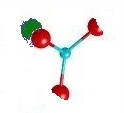} 
\centerline{\hskip 0.5cm a) Tetraborate and Tritetraborate \hskip 1.3cm b) Metaborate}
\caption{{\small Examples of positioning of an $Na^{+}$ ion}}
\label{fig:Tetraborates}
\end{figure}

We see that the rate of $B_4$ atoms steadily grows until $x$ reaches about $0.4$, then starts
to decrease. This is due to the fact that if throughout the process of consecutive modifier
addition the connectivity has to be maintained, i.e. no dangling bonds are being created. 
But even if all boron atoms were transformed into four-coordinated $B_4$ complexes, this will
saturate at $x = 0.5$. Beyond, another way of inserting the $Na^+$ ions takes progressively
place, the breaking of oxygen bonds between the borons and forming oxygens saturated with
$Na^+$, i.e. dangling oxygens. The boron atoms with such a saturated bond become in fact two-coordinate,
so that the overall connectivity remains maintained. The figure below displays two processes resulting 
from addition of the $Na_2O$ modifier:
\newline
a) Transforming a three-valenced boron into $B_4$ tetra-coordinate unit with an extra oxygen bond;
local connectivity increases 
\newline
b) Creating a non-bridging oxygen, i.e. breaking an oxygen bond; the connectivity decreases.

The result is measurable with NMR techniques; the relative amount of $4$-valenced borons is displayed on the left in Figure below.
It is noticeable that the glass transition temperature varies in a similar manner, as shown on the right. We discusse the dependence 
of $T_g$ on $x$ later on, in Sect. 6.

\begin{figure}[hbt]
\centering 
\includegraphics[width=7cm, height=4.4cm]{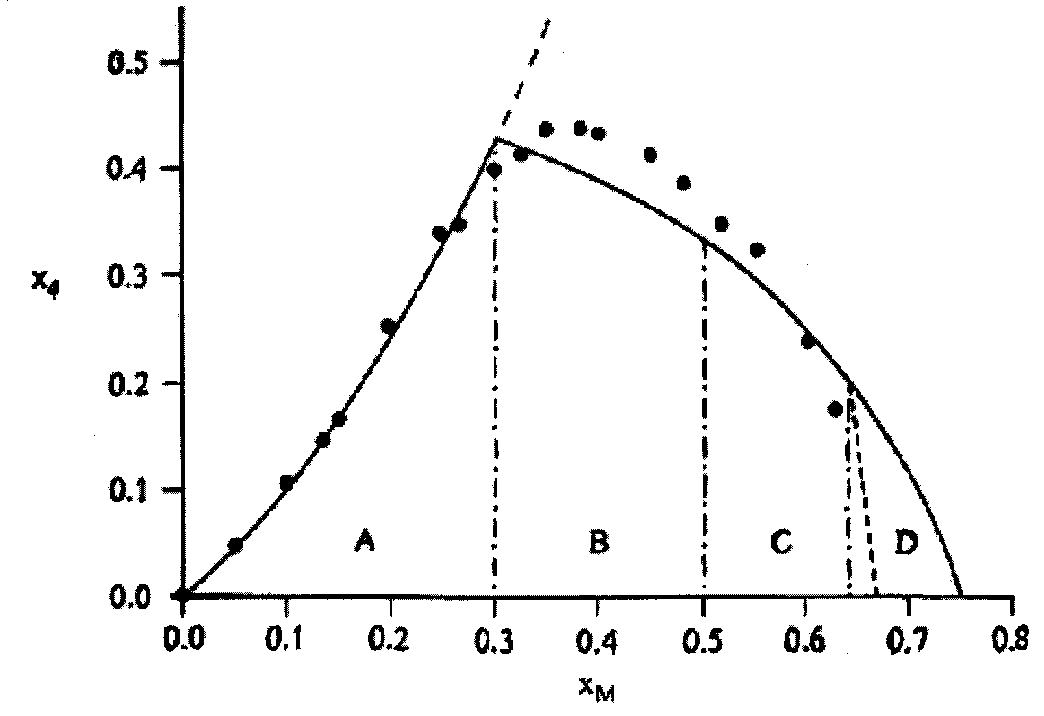}
\hskip 0.3cm
\includegraphics[width=6cm, height=4.8cm]{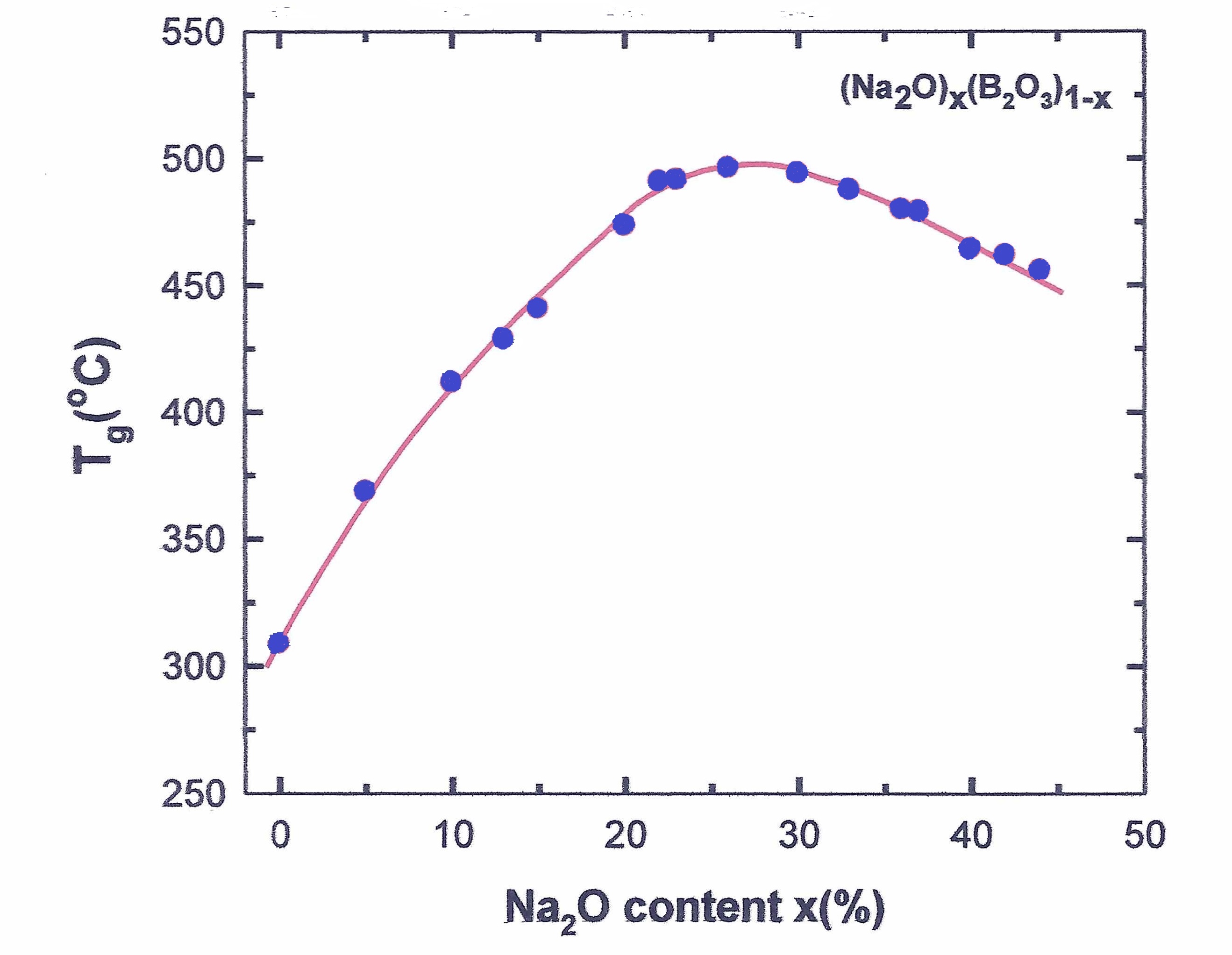}
\caption{\small{ Left: the rate of four-coordinate borons ($x_4$) with a $Na^+$ ion and an extra oxygen bond
as a function of modifier density $x_M$ (after A.Wright et al., 1997); Right: $(Na_2O)_{x} (B_2O_3)_{(1-x)}$ 
glass transition temperture $T_g$ as function of $x$ (after Vignobooran and Boolchand)) }}
\label{fig:Boronfour}
\end{figure}

We observe that the rate of $B_4$ atoms steadily grows until $x$ reaches about $0.4$, then starts
to decrease. This is due to the fact that if throughout the process of consecutive modifier
addition the connectivity has to be maintained, i.e. no dangling bonds are being created. 

Let us now proceed to a systematic display of clusters containing the $Na^+$ ions. 
We shall look firstly for clusters with four external bonds, so that they can replace the configurations $A$ and $B$ 
with no connectivity change in the network, 
and without creating dangling bonds.

The calculus of the average modifier content $x$ in each local configuration is easy, and will be given in each particular case. 

Besides, In each case, we shall proceed to the calculus of another
important parameter, the {\it rigidity} of each configuration displayed.
The latter is defined as the difference between the mean value of
the number of degrees of freedom per atom  $N_f$ and the ``free" value $3$,
$ \, r = N_f - 3$.

If  $r<0$, the cluster is  {\it floppy} or  {\it underconstrained};
\vskip 0.1cm
If $r=0$, the cluster is {\it isostatic};
\vskip 0.1cm
If $r>0$, the cluster is  {\it rigid} or  {\it overconstrained}.

The calculus of rigidity of given configuration is based on the
following assumptions concerning both {\it angular} and {\it stretching} constraints:
 
- Each three-coordinated boron atom, $B_3$ creates $3$ angular constraints (bond angles $= 120^0$);
\vskip 0.1cm
- Each tetra-coordinated boron atom, $B_4$, (with an extra oxygen bond) creates $5$ angular constraints;
\vskip 0.1cm
- An oxygen atom inside a boroxol ring creates $1$ angular constraint, whereas 
oxygen bonds out of rings have their angular constraints broken;
\vskip 0.1cm
- All covalent bonds without exception are equivalent to $1$ bond-stretching constraint;
finally, the $Na^+$ ions are not taken into account in the rigidity calculus, their position not being strictly defined.

As an example; let us evaluate the rigidity of clusters $A$ and $B$. In the first case (see \ref{fig:UnitsAB}) we have $4$ three-coordinate 
boron atoms, contributing $4 \times 3=12$ angular constraints, and three oxygens contained in the boroxol ring, thus contributing
$3 \times 1= 3$ angular constraints. The bridging oxygen relying the boroxol ring with the isolated $3$-coordinate boron is supposed
not to contribute to angular constraints. The linear (stretching) constraints are equal to the number of covalent bonds, here $12$,
 giving the total number of constraints equal $N_c=27$. The number of atoms involved in the $A$-configuration is $N_a = 10$ - four ``halves''
of bonding oxygens plus four borons and four oxygens. Dividing $N_c$ by $N_a$ we get the average number of constraints per atom
equal to $2.7$, which is $0.3$ short of $3.0$ which defines an {\it isostatic} configuration. Therefore the $A$-clusters are {\it floppy},
(or underconstrained), with $r = (N_c/N_a) - 3 = -0.3$.

 A similar calculus yields $N_c = 42, \; N_a = 15$, $r = (42/15) - 3 = -0.2$. The $B$-cluster is also floppy, albeit a bit less than
its $A$ counterpart.

In what follows, we use the classification and names of particular local clusters containing one or more
sodium ions as given by Wright {\it et al} in (\cite{Wright}).
Here are the structural units with one $Na^+$ ion, and with the same total connectivity
(coordination number $= 4$) as the original clusters $A$ and $B$:
\begin{figure}[hbt]
\centering 
\includegraphics[width=2cm, height=2.3cm]{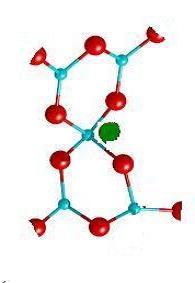}
\hskip 1.3cm
\includegraphics[width=2.1cm, height=1.8cm]{TritetraborateNa_halves.jpg}
\vskip 0.1cm
c) Pentaborate \hskip 0.9cm d) Tritetraborate 
\vskip 0.1cm
x= 0.167, \, \, r=0; \hskip 0.5cm x=0.25, \, \, r = 0. 
\caption{Two important clusters with one $Na^+$ ion; both are isostatic.}
\label{fig:Tripentabor}
\end{figure}
Adding more of $Na_2O$ leads to the multiplication of valences. To keep the connectivity balance, some $Na^+$ ions break oxygen bonds 
and remain in the vicinity of one of the oxygens.
\begin{figure}[hbt]
\centering 
\includegraphics[width=1.8cm, height=2.3cm]{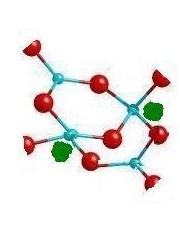}
\hskip 0.4cm
\includegraphics[width=2cm, height=2.4cm]{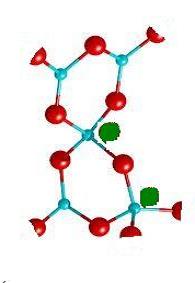}
\hskip 1.5cm
\includegraphics[width=2.8cm, height=1.7cm]{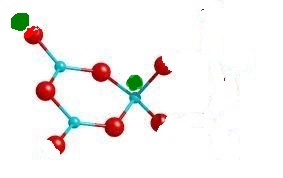} 
\hskip 0.1cm
\includegraphics[width=2.6cm, height=1.8cm]{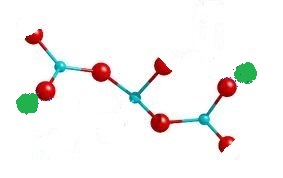} 
\vskip 0.2cm 
f) Diborate \hskip 0.4cm g) Dipentaborate \hskip 0.6cm h) Tetrametaborate \hskip 0.4cm e) Metaborate
\vskip 0.1cm
\hskip 0.4cm $x=0.333$ \, $r=0.176$ \hskip 0.3cm  $x= 0.286$ \,$r= 0.111$ 
 \hskip 0.2cm $x= 0.4$ \,  $r = - 0.176$
\caption{{\small Diborate (left) and two other clusters with two $Na^+$ ions. The total connectivity 
is still $5+3=8$}}
\label{fig:Dipentabor}
\end{figure}

The clusters with two $Na^+$ ions are: a {\it diborate} with connectivity $4$, and two new configurations with modified connectivity,
$5$ and $3$, {\it Di-pentaborate} and {\it Trimetaborate}. In order for connectivity being conserved, the $3$- and $5$-coordinate entities 
must appear in equal numbers. 
\hskip 0.2cm
\begin{center}
    \begin{tabular}{ | l | l | l | l | l | }
    \hline
    Name & Symbol & $B_4$-content & alkali $x_M$ & Rigidity $r$   \\ \hline \hline
    Single B3 & $$ \includegraphics[width=0.4cm, height=0.4cm]{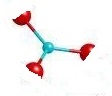}$$& 0 & 0 & -0.6  \\ \hline
    Single B4 & $$ \includegraphics[width=0.4cm, height=0.4cm]{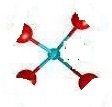}$$& 1 & 0.5 & 0  \\ \hline
    Boroxol ring & $$ \includegraphics[width=0.5cm, height=0.5cm]{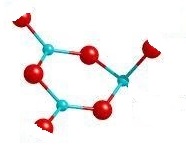}$$ & 0 & 0 & -0.2  \\ \hline
    A-cluster &  $$ \includegraphics[width=0.8cm, height=0.5cm]{ConfigA_halves.jpg}$$ & 0 & 0 & -0.3 \\ \hline
    B-cluster & $$ \includegraphics[width=0.8cm, height=0.5cm]{ConfigBhor_halves.jpg}$$   & 0 & 0 & -0.173  \\ \hline
    Pentaborate &  $$ \includegraphics[width=0.8cm, height=0.5cm]{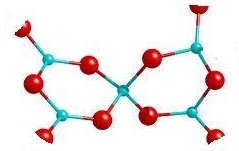}$$& 0.2 & +0.167 & 0 \\ \hline
    Tritetraborate & $$ \includegraphics[width=0.8cm, height=0.5cm]{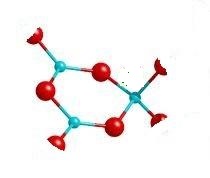}$$  & 0.333 & 0.25 & 0  \\ \hline
    Diborate & $$ \includegraphics[width=0.7cm, height=0.5cm]{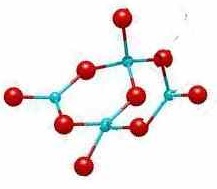}$$  & 0.5 & 0.333 & +0.182 \\ \hline
    Dipentaborate & $$ \includegraphics[width=0.7cm, height=0.5cm]{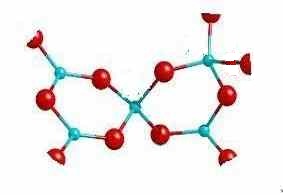}$$  & 0.4 & 0.286 & +0.111  \\ \hline
    Ditritetraborate & $$ \includegraphics[width=0.5cm, height=0.5cm]{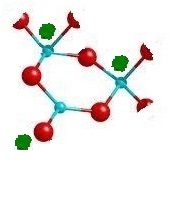}$$ & 0.667 & 0.4  & +0.176 \\ \hline    
    Metatriborate & $$ \includegraphics[width=0.5cm, height=0.5cm]{MetaTriborateNa_halves.jpg}$$ & 0.333 & 0.4  & -0.176 \\ \hline
    Metaborate & $$ \includegraphics[width=0.7cm, height=0.5cm]{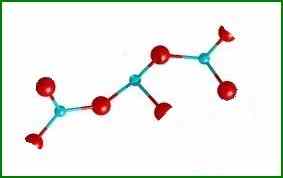}$$ & 0 & 0.4  & -0.882 \\ \hline
 \end{tabular}
\label{Clustersall}
\end{center}
{\centerline{{\small Table 1. Main local structures in alkali-borate glasses}}
\hskip 0.3cm
\indent
When the concentration of $Na_2O$ gets close to $40 \%$, new clusters are formed,
containing three $Na^+$ ions, with connectivity up to $6$ and the  the {\it pyroborate} clusters
containing three $Na^+$ ions, with connectivity $2$, but we shall not consider them here, restricting the range
of $x$ below the  $x=0.35$ limit.
Now we are able to define minimal sets of pure $B_2O_3$ clusters which can transform into new
clusters containing the $Na^+$ ions in such a way that the overall connectivity remains unchanged.

\section{Transforming clusters with alkali modifier}

According to the hypothesis exposed in Sect. $2$ and corroborated by numerous experimental data, a pure $B_2O_3$ melt about to undergo
a glass transition contains boroxol rings and isolated borons so that the amount of boron atoms trapped in $3$-fold boroxol rings
amounts to $83 \%$. After slow annealing, the resulting amorphous glassy network can be subdivided into $A$ and $B$ local clusters
each with exterior connectivity $4$, in proportion $2$ to $1$, i.e. $67 \%$ of $A$'s and $33 \%$ of $B$'s. 

Now suppose that a small amount of alkali modifier, say $Na_2O$ for example, is added to the melt, transforming a pure $B_2 O_3$ 
into a glassy network containing certain amount of four-coordinate borons $B_4$ due to the disruptive action of the $Na^{+}$ ions.
If the connectivity of the resulting network is to be mainained, this means that some number of local structures $A$ and $B$
had to be replaced by four-coordinate clusters containing $B_4$ borons created by the $Na^{+}$ ions. In what follows, we shall
consider the $Na_2O$ molecules dissolving into the melt and creating TWO local clusters with one $Na^{+}$ ion (i.e. with only one
$B_4$ tetraborate each), supposing that no clusters with two or more $Na^{+}$ ions can be spontaneously created at the onset of
the alkali oxide dissolving in the pure boron oxide melt. This requires in turn considering two local clusters at once, which
may be two $A$'s, an $A$ and a $B$, or two $B$'s; in each case, the total external connectivity is $2 \times 4 = 8$, and the
modified $Na^{+}$ containing clusters should display the same connectivity.

It should be made clear that these transformations of local structures should not be taken literally, like a genuine chemical
reaction. The two dissolved $Na^{+}$  and extra oxygen $O^{--}$ ions sneak their way into the network perhaps in a complicated
and chaotic manner, but at the end of the day the result can be expressed as a substitution of a pair of pure $B_2O_3$ clusters
 by $Na^{+}$ enriched ones, with strict connectivity conservation.  

While comparing many possible transformations of this sort, we shall try to minimize the rigidity variation and maximize the homogeneity
of the resulting local structures. 

Let us show how the molecules of $Na_2O$ can be inserted into the network with local connectivity remaining conserved.
The first ``reaction'' can be represented as follows:
\begin{figure}[hbt!]
\centering 
\includegraphics[width=1.4cm, height=2.5cm]{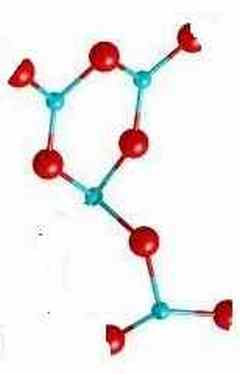}
\hskip 0.1cm
\includegraphics[width=1.4cm, height=2.5cm]{ConfigAver_halves.jpg}
\hskip 0.1cm
\includegraphics[width=1.5cm, height=1.9cm]{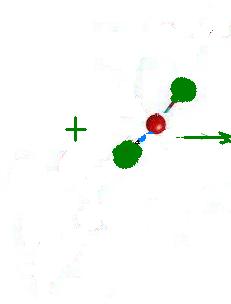}
\hskip 0.1cm
\includegraphics[width=1.5cm, height=2.8cm]{PentaborateNa_halves.jpg} 
\includegraphics[width=1.4cm, height=2cm]{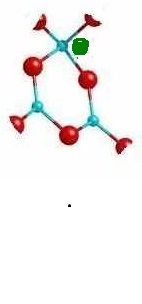}
\vskip 0.1cm
Two $A$ clusters \hskip 0.5cm  Triborate + Pentaborate
%
\caption{Insertion of one  $Na_2O$ molecule into a pure amorphous $B_2O_3$
network. The transformation can be encoded as $2 A + Na_2O \rightarrow P + T$. }
\label{fig:ReactionAA}
\end{figure}
Another reaction involving $A$ and $B$ clusters is shown below:
\begin{figure}[hbt!]
\centering 
\includegraphics[width=1.4cm, height=2.4cm]{ConfigAver_halves.jpg}
\hskip 0.1cm
\includegraphics[width=1.5cm, height=2.9cm]{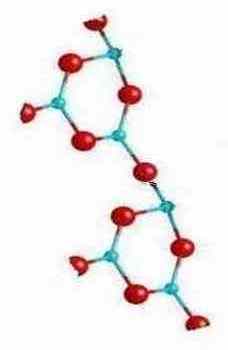}
\hskip 0.1cm
\includegraphics[width=1.5cm, height=1.9cm]{PlusNa2O.jpg}
\hskip 0.1cm
\includegraphics[width=1.5cm, height=2.8cm]{PentaborateNa_halves.jpg} 
\includegraphics[width=1.5cm, height=2.8cm]{PentaborateNa_halves.jpg}
\vskip 0.1cm
$A$ and $B$ clusters \hskip 0.5cm  Two pentaborates
%
\caption{Insertion of one  $Na_2O$ molecule into a pure amorphous $B_2O_3$
network. The transformation can be encoded as $A+B + Na_2O \rightarrow 2 P$. }
\label{fig:ReactionAB}
\end{figure}
Finally, there is another reaction involving one extra $Na_2O$ molecule  with local connectivity conservation,
involving two $B$ clusters:

\begin{figure}[hbt!]
\centering 
\includegraphics[width=1.5cm, height=2.7cm]{ConfigBver_halves.jpg}
\hskip 0.1cm
\includegraphics[width=1.5cm, height=2.7cm]{ConfigBver_halves.jpg}
\hskip 0.1cm
\includegraphics[width=1.5cm, height=1.9cm]{PlusNa2O.jpg}
\hskip 0.1cm
\includegraphics[width=3.2cm, height=2.4cm]{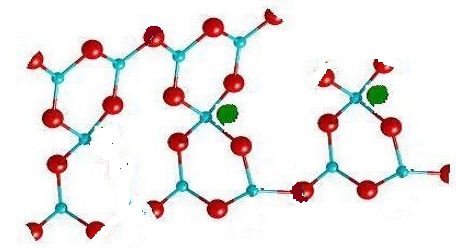} 
\vskip 0.1cm
\hskip 1.1cm Two $B$ clusters \hskip 1.5cm  A+Pentaborate+Tetratriborate
\caption{{\small Insertion of one $Na_2O$ molecule into a pure amorphous $B_2O_3$
network. This transformation can be encoded as $ 2 B + Na_2O \rightarrow P+T+A$. }}
\label{fig:ReactionBB}
\end{figure}

The three ``reactions'' shown in Figures (\ref{fig:ReactionAA}, \ref{fig:ReactionAB} and \ref{fig:ReactionBB}) above do in fact exhaust all possibilities
of insertion of a $Na_2 O$ molecule in the $B_2 O_3$ network with consequent disappearance of $A$ and $B$ clusters transformed into $P$'s and $T$'s
(pentaborates and tetratriborates), with a little amount of $A$ clusters stll remaining. 

At this point it is worthwhile to evaluate the relative production rate of pentaborates and tetratriborates, which are 
dominant at the first stages of alkali modifier's addition to the network. A given reaction's rate is proportional to the probability of 
picking up an appropriate couple of clusters, $A+A$, $A+B$ or $B+B$,
as shown in Figures (\ref{fig:ReactionAA}, \ref{fig:ReactionAB} and \ref{fig:ReactionBB}).
\begin{equation}
p_P \sim 1 \cdot p_A^2 + 2 \times 2 p_A p_B + 1\cdot p_B^2, \; \; 
p_T \sim 1 \cdot p_A^2 + 1 \cdot p_B^2, \; \; p_A \sim 1 \cdot p_B^2.
\label{probAPT}
\end{equation}
Inserting the initial rates of $A$ and $B$ clusters (i.e. ${\overset{(0)}{p}}_A=2/3, \; \; {\overset{(0)}{p}}_B =1/3$ ), we get the following linear approximation:
\begin{equation}
{\overset{(1)}{p}_P} \sim 
\frac{4}{9} +  \frac{8}{9} + \frac{1}{9}, \; \; \; {\overset{(1)}{p}_T} \sim \frac{4}{9} + \frac{1}{9}, \; \; \; {\overset{(1)}{p}_A} \sim \frac{1}{9}.
\label{probPTA1}
\end{equation} 
Next reactions resulting from inserting more $Na_2O$ molecules, with local connectivity conservation maintained:
\begin{figure}[hbt!]
\centering 
\includegraphics[width=1.2cm, height=2.3cm]{PentaborateNa_halves.jpg}
\hskip 0.1cm
\includegraphics[width=1.2cm, height=2.3cm]{PentaborateNa_halves.jpg}
\hskip 0.1cm
\includegraphics[width=1.5cm, height=1.9cm]{PlusNa2O.jpg}
\hskip 0.1cm
$\longrightarrow$
\hskip 0.1cm
\includegraphics[width=1.2cm, height=2.3cm]{DipentaborateNa_halves.jpg} 
\hskip 0.1cm
\includegraphics[width=1.2cm, height=2.3cm]{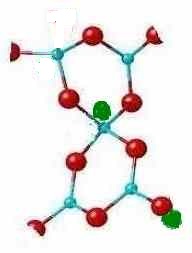}
\caption{{\small Insertion of one  $Na_2O$ molecule into a low alkali ($x \leq 18 \%$) amorphous $B_2O_3$
network. This transformation can be encoded as $2 P + Na_2O \rightarrow DP + MP$.}}
\label{fig:Reaction2P}
\end{figure}


\begin{figure}[hbt!]
\centering 
\includegraphics[width=1.5cm, height=2.4cm]{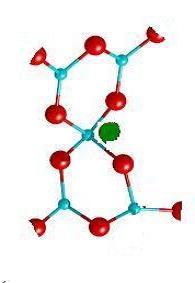}
\hskip 0.1cm
\includegraphics[width=1.4cm, height=1.9cm]{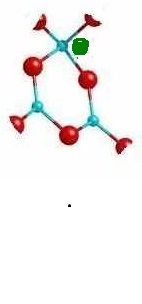}
\hskip 0.1cm
\includegraphics[width=1.5cm, height=1.9cm]{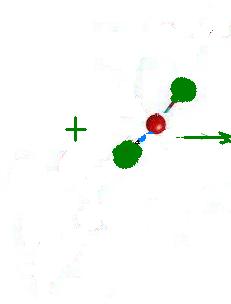}
\hskip 0.1cm
\includegraphics[width=1.6cm, height=1.6cm]{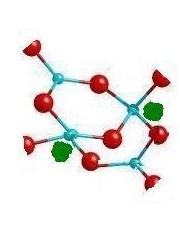} 
\includegraphics[width=1.6cm, height=1.6cm]{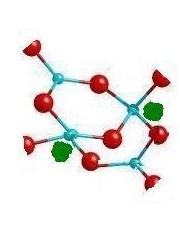}
\vskip 0.1cm
 Pentaborate + Tritetraborate \hskip 0.5cm  Two Diborates

\caption{Insertion of one  $Na_2O$ molecule into a pure amorphous $B_2O_3$
network. This transformation can be encoded as $P+T + Na_2O \rightarrow 2 D$.}
\label{fig:ReactionPT}
\end{figure}
An alternative issue of the same ``reaction'' can be also envisaged, producing new configurations,
a di-pentaborate and a meta-tetraborate, with respective connectivities $5$ and $3$ instead of
 $4$ plus $4$ in the case of two diborates:
\begin{figure}[hbt!]
\centering 
\includegraphics[width=1.4cm, height=2.4cm]{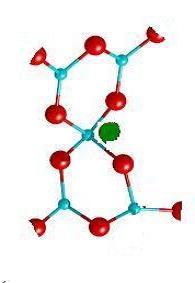}
\hskip 0.1cm
\includegraphics[width=1.4cm, height=2cm]{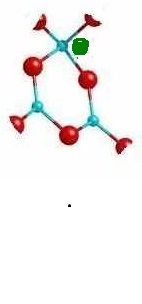}
\hskip 0.1cm
\includegraphics[width=1.4cm, height=1.9cm]{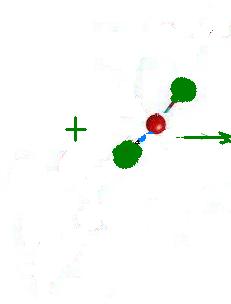}
\hskip 0.1cm
\includegraphics[width=1.5cm, height=2.1cm]{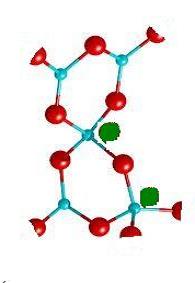} 
\includegraphics[width=2.4cm, height=1.7cm]{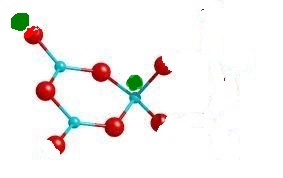}
\vskip 0.1cm
 Pentaborate + Tritetraborate \hskip 0.5cm  Dipentaborate and Tetrametaborate

\caption{Insertion of one  $Na_2O$ molecule into a low alkali content ($x\leq 18 \%$) borate glass.
network. This transformation can be encoded as $P+T + Na_2 O \rightarrow DP + MT$.}
\label{fig:ReactionDP}
\end{figure}

The Raman spectral analysis and NMR experiments have led to quite a precise picture
of evolution of numbers of various local configurations with continuous increase
of the $Na_2O$ modifier content.

The variation of abundance of local configurations is shown in the following figure:

\begin{figure}[hbt]
\centering 
\includegraphics[width=5cm, height=4.5cm]{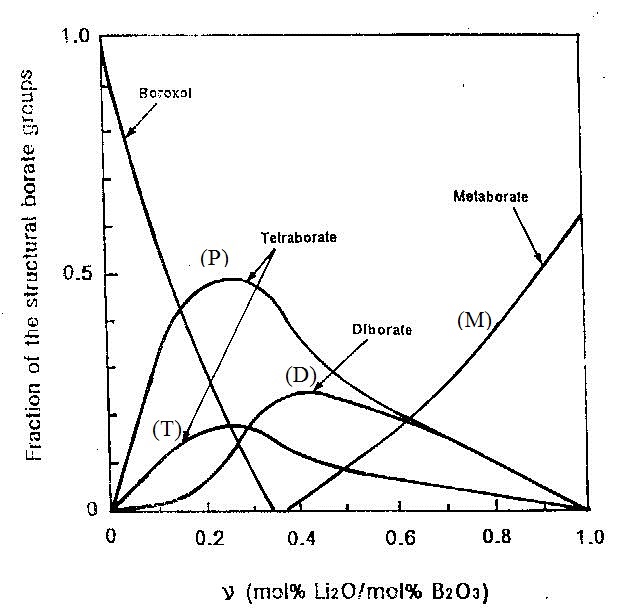}
\hskip 0.3cm
\includegraphics[width=5cm, height=4.5cm]{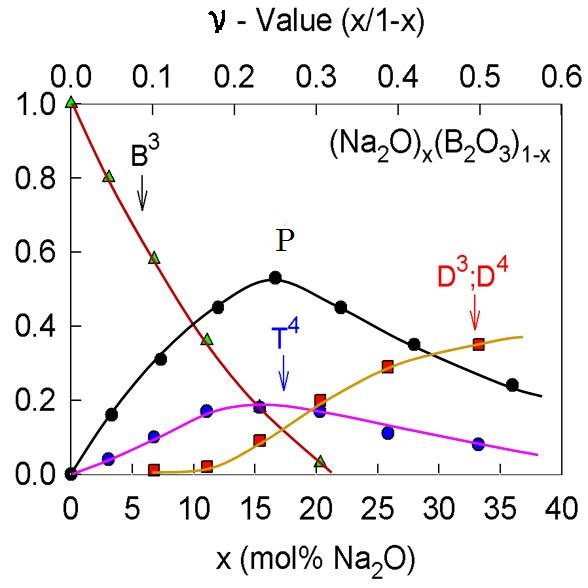}  
\caption{Abundance of main alkali-borate local configurations as function of modifier's molar density $x$.
Left: $(1-x) B_2O_3 + x Li_2O$, due to M. Balkanski and M. Massot; Right: $(1-x) B_2O_3 + x Na_2O$, courtesy of P. Boolchand. }
\label{fig:Borgroups1}
\end{figure}

\section{The Volterra approach}

The evolution of various "species" of boroxol clusters with progressive
addition of an alkali modifier is similar to the evolution of biological
systems with different living organisms, competing for food and space, or
even eating each other (predators and preys).

The simplest model is given by two species only, the prey $x$ and the 
predator $y$. The evolution of their (relative) numbers can be described as follows:

The prey population $x(t)$ increases at a rate $A x dt$,
proportional to its own number, but is simultaneously killed by predators at a rate $- B x y  dt$;

The predator population $y(t)$ decreases at a rate $  -C y dt$,
proportional to its own number, but increases at a rate $ D  x y  dt$;
which leads to the following differential system: 

$$ \frac{dx}{dt} = A x - B x y, \, \ \ \, \ \ \frac{dy}{dt} = - C y + D xy .$$

A typical solution of this system displays a neat quasi-periodic character: when rabbits proliferate, so do the foxes, which 
chase rabbits whose population decreases, thus condemning foxes to starve from hunger; this in turn gives the rabbits more
possibility to proliferate, and so on (see e.g. Kernerbook}

We shall apply similar method to the ``evolution" of abundances of various types of local
configurations (introduced in the perveious section and labeled $A, \, B, \, TT, \, P, \, D,$ etc.)
with groxing modifier concentration $x$, which variable will replace time in Lotka-Volterra
differential system. 

 The evolution of relative number of local configurations can be
described in a similar manner. The time parameter of the biological model is replaced
here by the modifier concentration $x$.

In order to establish the system of differential equations of Lotka-Volterra type,
let us analyze what happens to the network when a small amount of modifier,
 $\Delta n$ molecules, is added to the network, the constraint being connectivity conservation.

Let us add a small number of molecules $\Delta n$. to a network containing $N$ clusters of types 
$A$, $B$, $P$, $TT$, $D$, $M$ etc., their respective numbers being 
$N_A$, $N_B$, $N_P$, $N_{TT}$, $N_D$, $N_M$,, etc. 

In order to make the explanation of our model as clear as possible, let us
start with only first five configurations present,  $A$, $B$, $P$, $TT$ and $D$.
Notice that even if all local configurations ended up as being transformed into
diborates ($D$), the concentration $x$ could not bypass the mark $x=0.33$. Beyond
that concentration new local configurations appear, the Di-Tetraborates ($x = 0.4$),
 Tri-Pentaborates ($x = 0.375$), Di-Metaborates and Tri-Metaborates ($x = 0.5$), which we shall not take int
account in the simplified version of the model, valid only for $x$ below $0.35$ at most. 

Let us suppose then that at a given concentration $x$ of the $Na_2O$ alkali modifier,
 the sample glass obtained by annealing from melt contains $N$ molecules of the
glass former $B_2O_3$ and $n$ molecules of the modifier $Na_2O$. The relative concentration is
\begin{equation}
x = \frac{n}{N + n}.
\label{defx}
\end{equation}
Sometimes the ratio $\nu$ of $Na_2O$ molecules to the $B_2O_3$ molecules is used, and one has the obvious relation between 
these two parameters,
\begin{equation}
\nu = \frac{n}{N}, \; \; \; \nu = \frac{x}{1-x}, \; \; \; \; x = \frac{\nu}{1 + \nu}.
\label{defxnu}
\end{equation}
At the same time, the numbers of specific local configurations are:
$$N_A, \; N_B, \; N_{TT}, \; N_P \; \; {\rm and} \; \; N_D.$$
Let us denote the sum of all these numbers by $N_{tot}$; then
we can define the probabilities of finding at random one of the five local configurations as follows:
\begin{equation}
p_A = \frac{N_A}{N_{tot}}, \; \;  p_B = \frac{N_B}{N_{tot}}, \; \;  
p_{TT} = \frac{N_{TT}}{N_{tot}}, \; \; p_P = \frac{N_P}{N_{tot}}, \; \;  
 p_D = \frac{N_D}{N_{tot}}; 
\label{Probabs}
\end{equation}
Their sum is equal $1$ as it should be for the probabilities; this is why only four functions out of five are independent.
At the same time, the total number of $B_2O_2$ molecules in the sample can be found 
according to the obvious formula
\begin{equation}
N= 2 p_A + 3 p_B + \frac{3}{2} \, p_{TT} + \frac{5}{2} \, p_P + 2 p_D.
\label{Nmolecules}
\end{equation}  
In order to establish the differential system for the unknown functions 
$p_A (x)$, $ p_B (x)$, $ p_{TT} (x)$, $p_P (x) $ 
(the function $p_D (x)$ can be found then from the 
normalization relation (\ref{probAPT})), we must compare the numbers of the same configurations
after addition of some small amount of modifier in the form of  $\Delta n$ new molecules
of $Na_2O$ dissolved in the former melt and annealed to form a new glass. For this, we have
to decide what particular changes have occurred in the distribution of local clusters after
 the adjunction of new alkali molecules. It is enough to know the fate of one molecule
of $Na_2O$ dissociated in the melt, and evaluate the probablitities of various issues, then
multiplying these probabilities by $\Delta n$.

We have found the following reactions involving one $Na_2O$ molecule, leaving the total
connectivity of new clusters exactly the same as the initial ones. Here is the summary
of all such transformations leading to new local configurations:
$$ 2 A + Na_2O \rightarrow P + TT, \; \; \; \; \; \; A + B + Na_2O \rightarrow 2 P,$$
$$ 2 B + Na_2 O \rightarrow A + P + TT, $$
\begin{equation}
A + P + Na_2O \rightarrow 3 TT, \; \;  \; \; \; B + TT + Na_2 O \rightarrow P + D,
\label{Reactions}
\end{equation}
$$B + P + Na_2O \rightarrow A + TT + D, \; \; \; P + P + Na_2O \rightarrow 2 TT + D,$$
$$P + TT + Na_2O \rightarrow 2 D.$$
In the list of admissible connectivity-preserving transformations we have tacitly admitted
the principle according to which local configurations containing alkali ions are formed
progressively: the alkali-rich ones, containing two or more $Na^+$ ions, are formed only after 
most of the network has been transformed into configurations with one $Na^{+}$ ion only. As a matter
of fact, a reaction consisting in transformation of cluster $A$ with four boron atoms into
a diborate $D$ with four boron atoms and two $Na^+$ ions is theoretically possible, because it
also conserves the connectivity ($4$), but we consider its probability close to zero. 

The combination $TT + TT$ with addition of a molecule of $Na_2O$ can lead to configuration
conserving global connectivity only if one introduces new configurations with less than 
four external bonds, i.e. metaborates; at this stage we shall not count them in, stopping
the alkali concentration below $x = 0.3$, say. 

It follows from the above transformations (\ref{Reactions}) that a combination of one $Na_2O$ molecule
with two $A$-configurations transforms them in a pair $TT + P$ (a tri-tetraborate plus 
a pentaborate), i.e. one $Na_20$ molecule leads to the destruction of
two $A$ clusters and the creation of one $TT$ and one $P$ cluster. The resulting variation of
corresponding total numbers of each of the species taking part in the transformation is
$$\Delta N_A = - 2, \; \; \; \Delta N_{TT} = + 1, \; \; \; \Delta N_P = + 1.$$

The probability of such issue is proportional to the product of probabilities of picking
at random one $A$-cluster, i.e. $p_A^2$. Although there could be certain differences between
the energy barriers which may be different for particular transformations, at present stage we
shall not take them into account, assuming that the dominant feature and driving forces for 
transformations are the {\it connectivity} and {\it homogeneity} of the resulting network.
The result of the first reaction after the addition of $\Delta n$ alkali molecules is then
\begin{equation}
\Delta N_A \sim - 2 p_A^2 \, \Delta n, \; \; \Delta N_{TT} \sim + p_A^2 \, \Delta n, \; \; 
\Delta N_P \sim + p_A^2 \, \Delta n,
\label{firstdeltas}
\end{equation}
all other configuration numbers remaining unchanged by this reaction. Similarly, from the second
reaction of (\ref{Reactions}) we get the following variations:
\begin{equation}
\Delta N_A \sim - 2 p_A p_B \, \Delta n, \; \; \Delta N_B \sim - 2 p_A p_B \, \Delta n, \; \; 
\Delta N_P \sim + 4 p_A p_B \, \Delta n.
\label{seconddeltas}
\end{equation}
The probability of an encounter of two different configurations $A$ and $B$ is proportional
to $2 p_A p_B$; there is only one $A$ and one $B$ destroyed, but two $P$'s created, whence from
the factor $4$ in the last expression. Note also that in both cases the sum of all variations
is zero; this is because there are as many destroyed entities as the created ones, and the sum of
all probabilities remains normalized to one in the initial as well as in the final state.

The situation is a bit different with the third reaction of (\ref{Reactions}), because {\it three}
new configurations are created instead of the two ones  ($B + B$) that have disappeared. In order
to keep the sum of the contributions null, normalizing factor $2/3$ has to be introduced on the
right hand side, because we are comparing the initial probabilities related to {\it two} items 
with the final probabilities related to {\it three} new items. This yields the following account
of the result of the third reaction in (\ref{Reactions}):
{\small
\begin{equation}
\Delta N_A \sim \frac{2}{3}  p_B^2 \, \Delta n, \; \; \Delta N_B \sim - 2 p_B^2 \, \Delta n, \; \; 
\Delta N_{TT} \sim \frac{2}{3} p_B^2 \, \Delta n, \; \;
\Delta N_P \sim \frac{2}{3} p_B^2 \, \Delta n.
\label{thirddeltas}
\end{equation}}
With this in mind we can now evaluate and sum up the contributions coming from all the reactions
given in (\ref{Reactions}), arriving at the following result:
{\small
$$\Delta N_A = \left( - 2 p_A^2 - 2 p_A p_B - 2 p_A p_{TT} - 2 p_A p_P + \frac{2}{3} \, p_B^2 
+ \frac{2}{3} \, 2 p_B p_P \right) \, \Delta n,$$
$$ \Delta N_B = \left( - 2 p_A p_B - 2 p_B^2 - 2 p_B p_{TT} - 2 p_B p_P \right) \Delta n, $$
$$ \Delta N_{TT} = \left( p_A^2 + \frac{2}{3} (p_B^2 +  3 \times 2 p_A p_P +  2 p_B p_P + 2 p_P^2)
- 2 p_B p_{TT} - 2 p_P p_{TT} \right) \Delta n, $$
$$ \Delta N_P = \left( p_A^2 + 4 p_A p_B + \frac{2}{3} p_B^2 + 2p_B p_{TT}
 - 2 p_A p_P - 2 p_B p_P - 2 p_P^2 - 2 p_P p_{TT} \right) \Delta n, $$
\begin{equation}
\Delta N_D = \left( 2 p_A p_{TT} + 2 p_B p_{TT} + 4 p_P p_{TT} +
\frac{2}{3} ( 2 p_B p_P + p_P^2) \right) \Delta n.
\label{fivereactions}
\end{equation}}
One easily checks that the sum of all left-hand sides is zero, which means that only four of the
above five equations (\ref{fivereactions}) are linearly independent.

Before passing to  the continuous limit and form differential equations, let us express
everything exclusively in terms of probabilities and the unique independent variable, the
modifier's molecular content $x$. In order to change from $\Delta N=a$, $\Delta N_B$, etc., into $\Delta p_A$,
$\Delta p_B$, etc., it is enough just to divide both sides by the total number of configurations
As usually in differential calculus, the $\Delta N_i$
and $\Delta n$ are treated as infinitesimals, so there is no difference which actual values of $N_{tot}$
are chosen, the initial or the final ones.

Still, we have to express the ratio $\Delta n/N_{tot}$ in terms of the differential $\Delta x$.
At the moment, on the right-hand side we have got the ratio $\Delta n/ N_{total}$; but this can be
easily transformed into the quantity $\Delta \nu$ as follows. Remember that there is a simple
relationship between $N_{tot}$ given in (\ref{Nmolecules}); therefore,
dividing both sides by $N_{tot}$; we get, by definition of configuration probabilities,
\begin{equation}
\frac{N}{N_{tot}} = 2 p_A + 3 p_B + \frac{3}{2} p_{TT} + \frac{5}{2} p_P + 2 p_D = < k >,
\label{TwoNs}
\end{equation}
where we note by $<k>$ the {\it average number of $B_2O_3$ molecules} per local configuration, so that one can write
\begin{equation}
N = <k> N_{tot}, \; \; \; {\rm so  \; that} \; \; \frac{\Delta n}{N_{tot}} =
<k> \frac{\Delta n}{N} = <k> \, \Delta \nu.
\label{simpletotal}
\end{equation} 
Now we can proceed to the continuous limit, dividing by $\Delta \nu$ both sides of the equations
(\ref{fivereactions}). As an example, let us write down the first of the five equations:
\begin{equation}
\frac{d p_A}{d \nu} = <k> \left( - 2 p_A^2 - 2 p_A p_B - 2 p_A p_{TT} - 2 p_A p_P + \frac{2}{3} \, p_B^2 
+ \frac{2}{3} \, 2 p_B p_P \right),
\label{firstdiffnu}
\end{equation}   
and we remind that $<k> = 2 p_A + 3 p_B + \frac{3}{2} p_{TT} + \frac{5}{2} p_P + 2 p_D$.
The derivation with respect to the variable $\nu$ can be transformed into the derivation with respect
to the variable $x$ using the relation (\ref{defxnu}); we have
\begin{equation}
\frac{d \;}{d \nu} = \frac{d x}{d \nu} \, \frac{d \;}{dx} = \left[ \frac{d \nu}{d x} \right]^{-1}
\, \frac{d \;}{dx} = (1-x)^2 \, \frac{d \;}{dx}.
\label{reldiffxnu}
\end{equation}
Now the first differential equation of (\ref{fivereactions}) can be written as follows:
\begin{equation}
\frac{d p_A}{dx} = \frac{<k>}{(1-x)^2} 
\left( - 2 p_A^2 - 2 p_A p_B - 2 p_A p_{TT} - 2 p_A p_P + \frac{2}{3} \, p_B^2 
+ \frac{2}{3} \, 2 p_B p_P \right),
\label{oneoffive}
\end{equation}
and similarly for the four remaning equations, which are constructed in the same manner as (\ref{oneoffive}), by replacing
the right-hand side expression in the parentheses by corresponding expressions appearing in equations (\ref{fivereactions}).

Let us consider the simplified version of the system
valid at the onset of modificator's addition, for low values of $x \leq 0.25$.
In this case, the system can be linearized and solved almost immediately. The initial
conditions are clear: at $x=0$ we have
$$p_A (0) = \frac{2}{3}, \; p_B (0) = \frac{1}{3}, \; 
p_{TT} (0) = 0, \; p_P (0) = 0, \; p_D (0) = 0.$$
It is also obvious that at the very beginning, the dependence of $p_{TT}, \; p_P$ on
$x$ is linear in $x$, while $p_D$ can be only quadratic in $x$. 
Keeping only the powers of $p_A$ and $p_B$ and neglecting the $p_{TT}, \; p_P$ and $p_D$
in the right-hand sides leads to the following approximate system at $x$ close to $0$:
$$\frac{d p_A}{dx} \simeq \left( - 2 p_A^2 - 2 p_A p_B  + \frac{2}{3} p_B^2 .. \right)
\left( 2 p_A + 3 p_B ...  \right),$$
$$\frac{d p_B}{dx} \simeq \left( - 2 p_B^2 - 2 p_A p_B  .. \right)
\left( 2 p_A + 3 p_B ...  \right),$$
$$\frac{d p_{TT}}{dx} \simeq \left( p_A^2 + \frac{2}{3} p_B^2 .. \right)
\left( 2 p_A + 3 p_B ...  \right),$$
\begin{equation}
\frac{dp_P}{dx} \simeq \left( p_A^2 + 4 p_A p_B + \frac{2}{3} p_B^2 .. \right)
\left( 2 p_A + 3 p_B ...  \right),
\label{fiveapprox}
\end{equation}
We do not write down the fifth equation (for $p_D (x)$, because at this stage $p_D$
is of the order of $x^2$, therefore can be neglected.
A further simplification can be made by taking into account the constant ratio $p_B = p_A/2$
which should remain valid for very small amounts of alkali modifier and the fact that the rate of other
reactions involving pairs of $P$, $T$ or $D$ are negligeable until the $A$ and $B$ configurations prevail,
 which remains true up to $X= 0.2$.

Therefore at zeroth 
approximation we can replace $p_B$ by $p_A/2$ and keep only the constant terms on the right-hand 
side of the differential equations (\ref{fivereactions}). This yields the following approximate 
system in which all but constant terms have been neglected, $p_A$ replaced by its initial value
$2/3$ and $p_B$ by its initial value $1/3$:
$$\frac{d p_A}{dx} = - \frac{119}{12} p_A^3, \; \; \frac{d p_B}{dx} = - \frac{63}{12} p_A^3, \; \; 
\frac{d p_{TT}}{dx} = \frac{49}{12} p_A^3, \; \; \frac{d p_P}{dx} = \frac{133}{12} p_A^3. $$
The fractions on the right-hand sides are so close to integer numbers, that we shall use the
following approximation:
\begin{equation}
\frac{d p_A}{dx} = - 10 p_A^3, \; \; \frac{d p_B}{dx} = - 5 p_A^3, \; \; 
\frac{d p_{TT}}{dx} = 4 p_A^3, \; \; \frac{d p_P}{dx} = 11 p_A^3, \; \; 
\label{fiveintegers}
\end{equation}
It is easy to solve the first equation by direct integration, which gives
\begin{equation}
\frac{d p_A}{p_A^3} = - 10 dx \rightarrow p_A^{-2} = 20 x + C,
\label{pAsolution}
\end{equation}
C being the integration constant. At $x = 0$ the value of $p_A^{-2}$ is $9/4$, so $C = 9/4$,
and we get the solution 
\begin{equation}
p_A (x) = \frac{2}{3} \, \left( 1 + \frac{80}{9} x \right)^{- \frac{1}{2}}.
\label{pAsqrt}
\end{equation}
Expanding (\ref{pAsqrt} around $x=0$ we get the following approximate
solution for $p_A(x)$ and $p_B(x) = p_A(x)/2$ in the vicinity of $x=0$:
\begin{equation}
p_A(x) \simeq \frac{2}{3} - \frac{80}{27} \, x, \; \; \;
 p_B(x) \simeq \frac{1}{3} - \frac{40}{27} \, x,
\label{fivefractions}
\end{equation}
Keeping only the constant terms on the right-hand sides of two subsequent equations we get easily
$$\frac{d p_{TT}}{dx} \simeq 4 p_A^3 = \frac{96}{81}, \; \; \; 
\frac{d p_P}{dx} \simeq 11 p_A^3 = \frac{266}{81} \rightarrow  p_{TT} \simeq \frac{96}{81} \, x, \; \; \; 
p_P \simeq \frac{266}{81} \, x.$$
At this approximation stage the diborates are still ``invisible", because their number is
proportional to $x^2$ and does not appear in linear approximation.

Even at this stage of very crude approximation we can get predictions concerning the
derivatives of abundance curves displayed in figure (\ref{fig:Borgroups1}) at $x=0$.
According to the approximate linear solutions, the derivatives with respect to $x$
at $x=0$ take on the following values:
\begin{equation}
\frac{d p_A}{dx} (0) \simeq - 3, \; \; \frac{d p_B}{dx} (0) \simeq -1.5, \; \; 
\frac{d p_{TT}}{dx} (0) \simeq 1.2,  \; \; \frac{d p_P}{dx} (0) \simeq 3.3,
\label{valuesatzero}
\end{equation} 
where we have used approximate values of fractions appearing in (\ref{fivefractions}). 
The abundance of boroxol rings is evaluated as $\frac{3}{4} p_A + p_B$ this yields the
value of derivative at $x=0$ of the curve representing boroxol rings' abundance being
equal to $- 3.85$.

\begin{figure}[hbt!]
\centering 
\includegraphics[width=6cm, height=4.3cm]{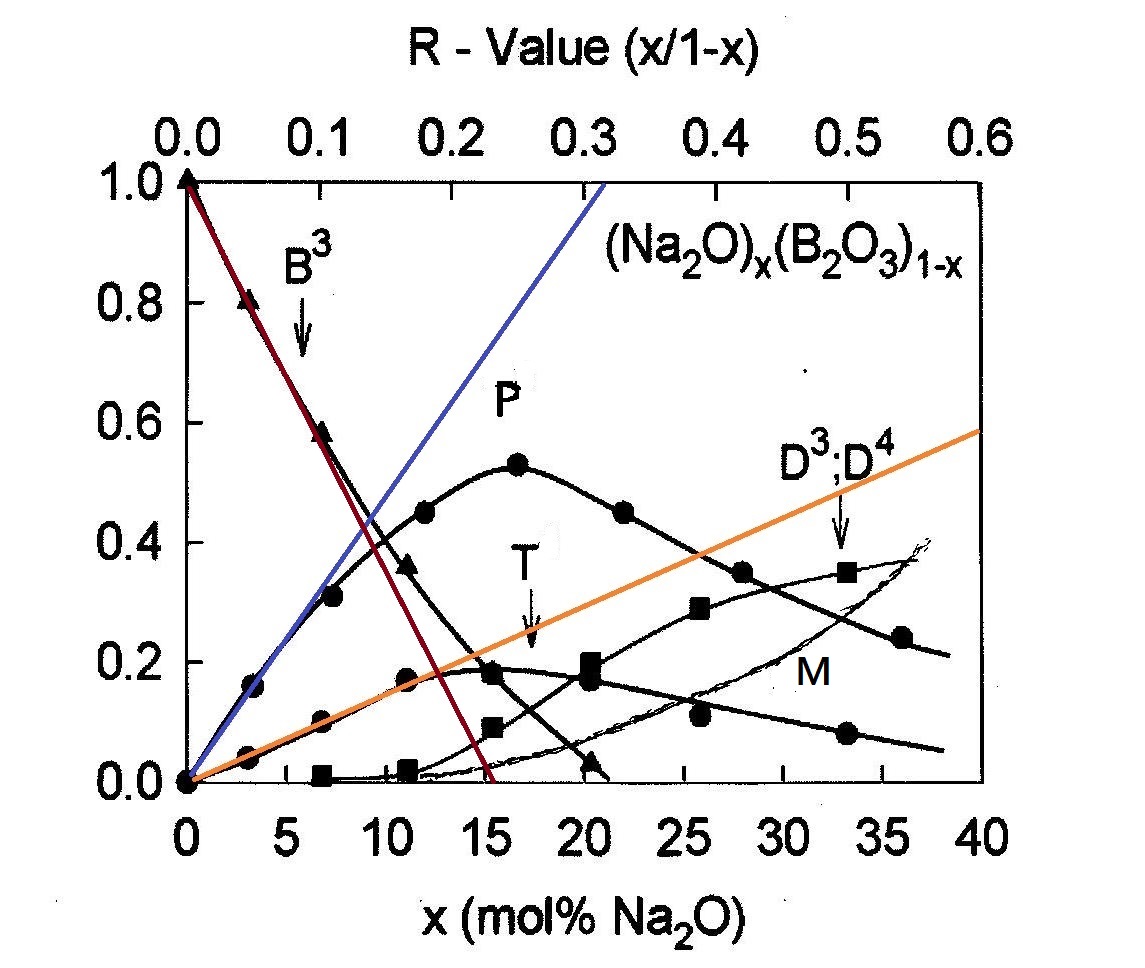}
\hskip 0.3cm
\includegraphics[width=5cm, height=4.2cm]{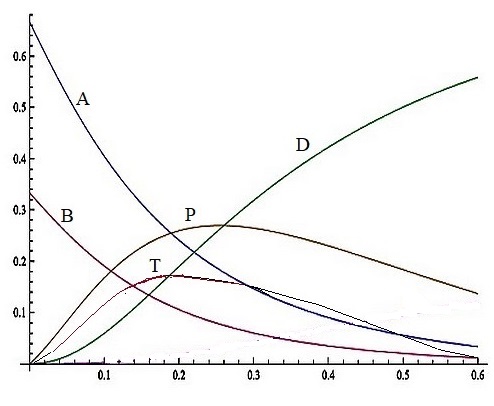}
\caption{{\small{Left: The plot showing the relative abundance of local configurations $P, T, D$ and $M$ as well as the pure borate clusters $A$ and $B$ 
with linear approximations at $x=0$ shown with straight lines; Right: solutions of the first approximation system obtained with Maple programme. They are
close enough to the experimental curves only at very low alkali concentrations ($x \leq 0.1$)}}}
\label{fig:Borolines}
\end{figure}

Comparing these values with the curves in (\ref{fig:Borgroups1}), we see that they are in a quite 
fair agreement with the experimental data.  

\section{Rigidity and glass transition temperature}

The glass transition temperature in covalent glasses depends crucially on topological properties of random network, in first place on its {\it connectivity}. 
The simplest and most compact expression of this complex statistical feature is the {\it average coordination number} (\cite{RKJNCS}, \cite{Kerner1995}, \cite{Kernerbook}),
which represents a purely topological characteristic of a random network, and contains no information about forces and energies involved. Its influence
on the glass transition temperature is of exclusively entropic nature. A simple rula was derived expressing the initial slope of the curve $T_g(x)$ at
working very well in covalent binary chalcogenide glasses like $Se_x As_{(1-x)}$ or $As_x Ge_{(1-x)}$. If the primary glass former's atoms are $m$-valenced and
the modifier's atoms are $m'$-valenced, the derivative of the glass transition temperature curve is given by the following formula:
\begin{equation}
\frac{d T_g}{d x}\mid_{x=0} = \frac{T_g (x=0)}{\ln(\frac{m'}{m})}.
\label{MagicTg}
\end{equation}
It works perfectly well for covalent random network glasses, but much less so for oxides, in particular borates and silicates, 
displaying local ring structures. 

 A simple model of glassy thermodynamics by G.G. Naumis(\cite{Naumis2006}) relates
the glass transition temperature $T_g$ with number of {\it floppy modes} in a given glass.
The formula relating $T_g$ with the density $f$ of floppy modes has the following form:
\begin{equation}
T_g(f) = \frac{T_g (f=0)}{1 + \alpha \, f}.
\label{floppymodes}
\end{equation}
The density of ``floppy modes'' among all vibrational modes in a given glass network, although not identical with ``zero frequency modes'',
can be directly related to the density of broken angular constraints of bridging oxygens non involved in boroxol rings. 

The formula (\ref{floppymodes}) can be compared with the Gibbs-Di Marzio phenomenological formula using the average coordination number $c$: 
\begin{equation}
T_g(c) = \frac{T_g(<c> = 2)}{1 - \beta(<c> - 2)}, \; \; \; \; \beta = \frac{5 \alpha}{2 \alpha + 6}.
\label{GibbsdiMarzio}
\end{equation}
In covalent glasses, the Gibbs-Di Marzio formula applies with the value of $\beta = 0.72$.
We may try a similar formula as function of our rigidity defect parameter $<r>$ as follows:
\begin{equation}
T_g(<r>) = \frac{T_g^{iso}}{1 - \gamma <r>}
\label{Tgrigid}
\end{equation}
The best fit for our version of $T_g$ for alkali-borate glass is when $\gamma = 1.45$, close to $2 \beta$ of the Gibbs-Di Marzio formula.%


At a given value of the modifier concentration $x$ we can evaluate the
The average rigidity defect of the network at a given value of the modifier concentration $x$ can be evaluated taking the mean statistical 
value of $r$:
\begin{equation}
<r> = {\displaystyle{\sum_{i} }}  p_i \, r_k,  \; \; {\rm with} \; \;    i=  A, \,  B, \,  P, \,  TT, \,  D, etc.
\label{meanr}
\end{equation} 
The values of $r$ for particular configurations were given previously. The average rigidity defect as a function of $x$ are given in Table II below,
followed by the curve $T_g (x)$ resulting from the dependence of $T_g$ on $<r> =<r> (x)$ given by the formula (\ref{Tgrigid}).

%
%
%

{\small
\begin{center}
\begin{tabular}{|c|c|c|c|c|c|c|c|}
\hline
\raisebox{0mm}[4mm][2mm]{ $x$ } & $p_A$ & $p_B$ & $p_P$ & $p_T$ & $p_D$ & $p_M$& $<r>$ \cr
\hline\hline
\raisebox{0mm}[4mm][2mm]{ $0.00$ } & $0.67$ & $0.33$ & $0$ & $0$ & $0$ & $0$ & $-0.257$ \cr
\hline
\raisebox{0mm}[4mm][2mm]{ $0.08$ } & $0.42$ & $0.21$ & $0.28$ & $0.19$&$0$ & $0$ &$- 0.161$  \cr
\hline
\raisebox{0mm}[4mm][2mm]{ $0.15$ } & $0.267$ & $0.136$ & $0.43$ & $0.17$ & $0$ & $0$ &$-0.104$ \cr
\hline
\raisebox{0mm}[4mm][2mm]{ $0.20$ } & $0.167$ & $0.083$ & $0.52$ & $0.16$ & $0.07$ & $0.04$ &$-0.082$ \cr
\hline
\raisebox{0mm}[4mm][2mm]{ $0.25$ } & $0.107$ & $0.05$ & $0.53$ & $0.19$ & $0.09$ & $0.05$ & -0.045 \cr
\hline
\raisebox{0mm}[4mm][2mm]{ $0.30$ } & $0.033$ & $0.017$ & $0.46$ & $0.16$& $0.17$ & $0.08$ & -0.094 \cr
\hline
\raisebox{0mm}[4mm][2mm]{ $0.35$ } & $0$ & $0$ & $0.43$ & $0.17$& $0.22$& $0.18$ & -0.104\cr
\hline
\raisebox{0mm}[4mm][2mm]{ $0.40$ } & $0$ & $0$ & $0.38$ & $0.16$& $0.28$& $0.19$ & -0.108\cr
\hline \hline
\end{tabular} 
\end{center}}
\centerline{Table II: \,\small{Average rigidity defect $<r>$ for different concentrations of $Na_2O$.}}
\centerline{\small  The values calculated
averaging from experimental data in (\ref{fig:Borgroups1}).}

We observe that alkali borate glasses $(Na_2O)_x (B_2O_3)_{(1-x)}$ are {\it floppy} from $x \leq 0.2$, close to {\it isostatic}
in the range $0.2 < x < 0.3$ and floppy again beyond $x>0.3$.

%
%
%
%

\begin{figure}[hbt]
\centering 
\includegraphics[width=8.5cm, height=5cm]{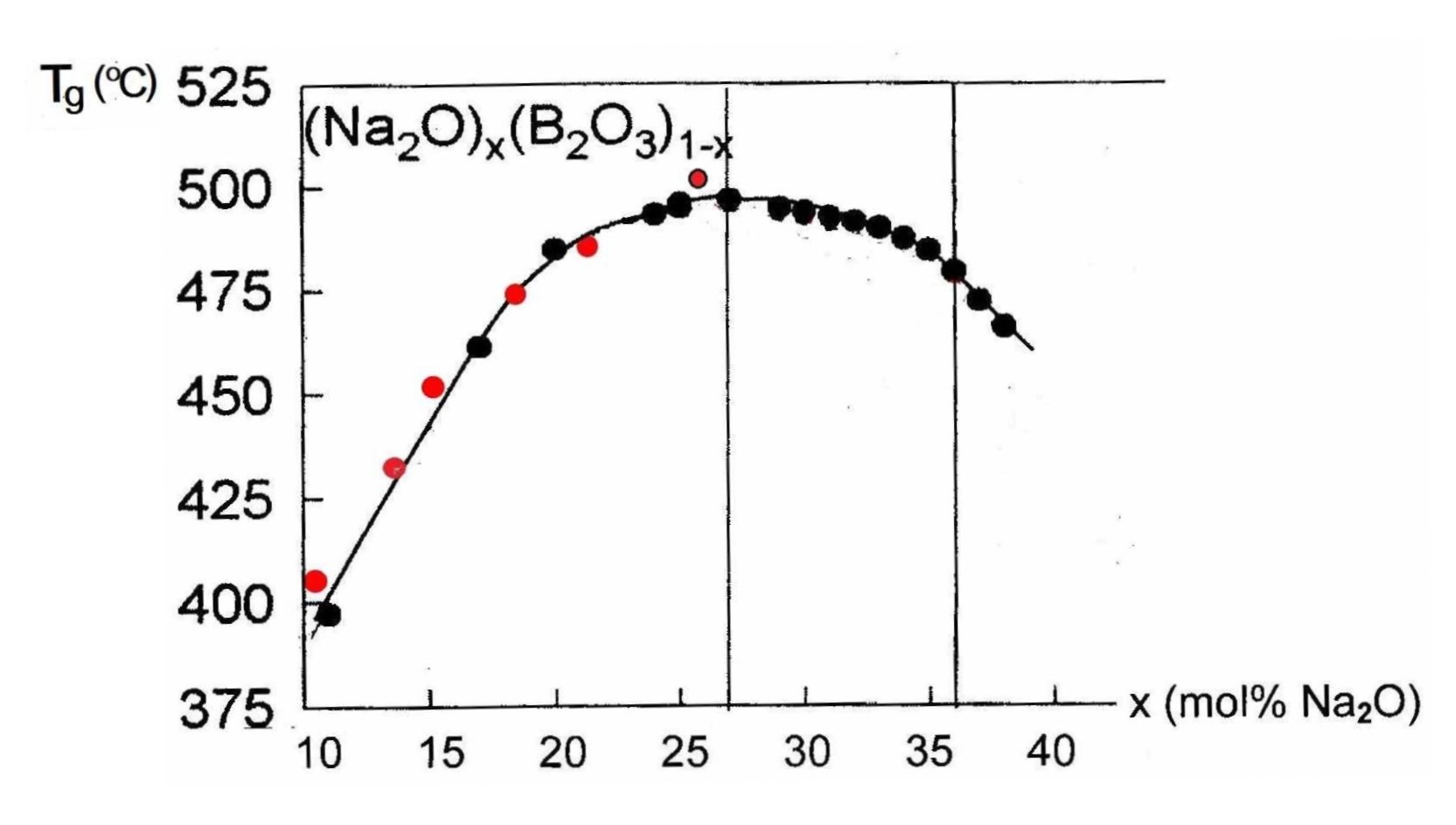}
\caption{{\small The glass transition temperature $T_g$ versus $Na_2O$ molar concentration.
Black: Experiment (courtesy P. Boolchand), Red: theory (formula (\ref{Tgrigid}) }}
\label{fig:Glasstemp}
\end{figure}

\indent
\hskip 0.5cm
{\bf Acknowledgements}
\vskip 0.3cm
\indent
One of us (R.K.) gratefully acknowledges many inspiring and fruitful discussions with Rafael Barrio, James C. Phillips, Punit Boolchand
and Matthieu Micoulaut. Special thanks are due to Punit Boolchand for generous sharing his experimental results.

\end{document}